# Spasers, VCSELs, and Surface plasmon emitting diodes (SPED's): their unique features and figures of merit.

Jacob B. Khurgin

Department of Electrical and Computer Engineering, Johns Hopkins University

Baltimore, Maryland 21218

Greg Sun

Department of Physics, University of Massachusetts Boston

Boston, Massachusetts 02125

**Abstract**

We compare fundamental characteristics of proposed electrically pumped subwavelength plasmonic lasers (spasers) with ubiquitous all-dielectric semiconductor micro-lasers (VCSEL's) and indicate their close similarities as well as large differences in performance (threshold, efficiency, coherence and speed). These results should assist researchers in making informed choice of the emitter for each particular opto-electronic application.

**Introduction**

Recent years have seen a burst of activity in nanophotonics directed towards development of functional integrated optical circuits capable of if not supplanting integrated electronics circuits of today, then at the very least augmenting them. Significant progress has been achieved in the development of miniature passive (waveguides, couplers, diffractive elements) and active (modulators and lasers) components and first integrated circuits are already reaching commercialization stage. One of the main obstacles to further increasing the density of integration has been and remains the diffraction limit that precludes shrinking the device dimensions below roughly half of wavelength. The diffraction limit is insurmountable in the all-dielectric (including semiconductor) optical devices, yet it can be circumvented when the metal is introduced. This fact has given the impetus to a new field of nanoplasmonics with the goal of shrinking the dimensions of optoelectronic devices, passive and active, beyond the optical wavelength [1,2]. Unfortunately, the oscillations of free carriers in the metal (surface plasmons–SP) that facilitate sub-wavelength confinement of light are strongly damped, primarily due to the large density of states available for the scattering of carriers, and, as shown in [3,4] the rate of loss of energy for any sub-wavelength plasmonic element is on the order of 10 fs$^{-1}$, many orders of magnitude above the typical loss rate in dielectrics and undoped semiconductors, and about two orders of magnitude higher than in doped semiconductors. High loss has impeded the progress towards practical nanoplasmonic devices and brought the issue of compensating this high loss with gain to the forefront, but practical loss compensation has not been achieved yet, and as our prior research has shown [5] full loss compensation may only be possible in the structures where the confinement is not much better than in all dielectric devices (so-called long range plasmons). It was shown that Purcell enhancement of spontaneous emission was the culprit behind the high pumping density required to compensate the loss in plasmonic devices which are significantly smaller than wavelength of light in all three dimensions.

Closely related to the issue of loss compensation with gain is that of development of sub-wavelength coherent source – essentially if the gain can be large enough to compensate both loss in the metal and radiative (or coupling) loss of photons escaping the cavity then one can achieve coherent oscillations of confined surface plasmons coupled with photons (localized surface plasmon polaritons- LSPP). The new class of devices is alternatively referred to as a plasmonic nanolaser [6], or, more intriguingly, a spaser [7,8]. There have been a number of demonstrations of plasmon-assisted nanolasers, both optically [9] and electrically [6] pumped, but these devices were invariably significantly longer than a wavelength in dielectric and the threshold was also invariably high. When it comes to true subwavelength spaser in all three dimensions there has been but one optically-pumped demonstration, also with a very high threshold [10]. In our previous work [11,12] we had shown that for the electrically pumped spaser threshold current density will always exceed MA/cm$^2$, the fact that has been acknowledged by most of the community [13]. Whether such current densities (three orders of magnitude higher than in a typical semiconductor laser) are practical remains an open question, but in this work we would like to examine the other issue, regarding the sub-wavelength spasers, namely whether the spaser properties are so unique and promising revolutionary change in nanophotonics that their development must be pursued at all cost.

Among the properties of spasers that are considered revolutionary besides the obvious integration density are the fact that they are fundamentally different from lasers in a sense that they do not emit electro-magnetic waves [8], their low power consumption and of course high speed, reaching into THz. In this work we scrutinize these claims one by one by comparing spasers with the state of the art ubiquitous miniature vertical cavity surface emitting semiconductor lasers – VCSEL's [14,15] and conclude that there are no fundamental differences between the two except for the fact that the threshold power dissipation in a spaser is higher. We also show that spaser operating above threshold is not superior to the incoherent source of SPP's – surface plasmon emitting diodes (SPED) that require many orders of magnitude smaller pump power.

**Is spaser fundamentally different from a small VCSEL?**

According to the spaser description in the literature [8], the main difference between the spaser and laser consists of the fact that while in the conventional laser the energy is emitted in the form of coherent photons, i.e. the energy alternates between the energies of electric and magnetic fields, in the a spaser the photons are replaced by surface plasmons i.e. the combination of charge carrier oscillations and electric fields. However, it should be noted that the difference is far less categorical than that. In fact, in both cases, it is **polaritons** that are created. In spaser the quantum of energy is a **surface plasmon polariton**, i.e. coupled oscillations of **free** electron charge in the metal and the electric field, with a small amount of magnetic field. In any solid state laser (including semiconductor) the quantum of energy is also a **polariton,** i.e. coupled oscillations of electric and magnetic fields coupled with oscillations of **bound** electrons in the dielectric (semiconductor). The difference is mostly qualitative and to ascertain it we shall calculate the fraction of the energy of the emitted quantum that is contained in the field oscillations $f_{field}$ and the remaining fraction $1-f_{field}$ that is contained in the carrier oscillations.

Consider a semiconductor laser operating with say GaAs VCSEL (Fig. 1a) whose dielectric constant is $\varepsilon_r = 1 + \chi$ where $\chi$ is optical susceptibility. The total energy density in the lasing mode inside the cavity is

$$U = \frac{\varepsilon_0}{4} \frac{\partial(\varepsilon_r \omega)}{\partial \omega} E^2 + \frac{\mu_0}{4} H^2 = \frac{\varepsilon_0 E^2}{4} \left[ 1 + \frac{\partial(\chi \omega)}{\partial \omega} + \varepsilon_r \right] \quad (1)$$

where the first term in the square brackets is the energy of electric field, the third one (obtained using the relation $H = E \varepsilon_0 \varepsilon_r / \mu_0$ is that of magnetic field, and the second term is the sum of kinetic and potential energies of oscillating bound electrons (and to a smaller degree ions) . It is clear that this term becomes large around the vicinity of resonance in a dispersive medium (or close to the bandgap) where in addition to the potential energy the kinetic energy of oscillating bound electrons becomes large. Taking the data for GaAs at 870 nm one obtains $\partial(\chi \omega)/\partial \omega \approx 21$ and $\varepsilon_r \approx 13$ resulting in

$$f_{field}^{VCSEL} = \frac{1 + \varepsilon_r}{1 + \varepsilon_r + \partial(\chi \omega)/\partial \omega} \approx 0.4 \quad (2)$$

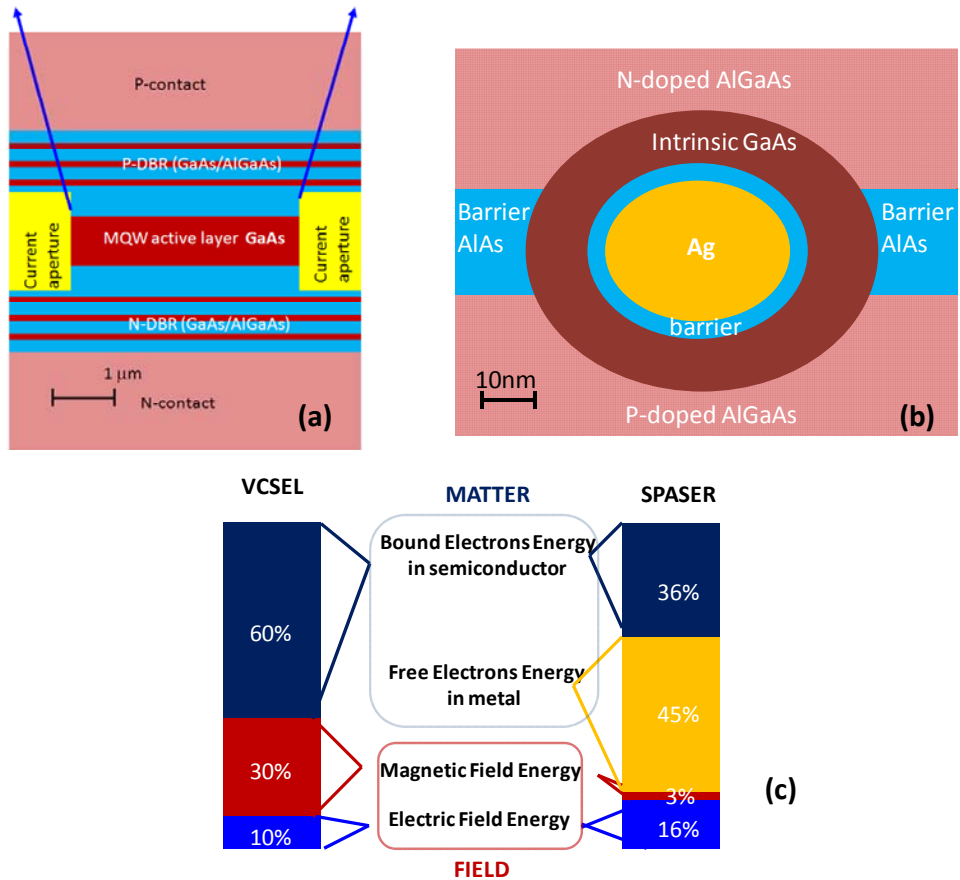

**Figure 1.** Sketches of (a) VCSEL, (b) Electrically-pumped spaser, and (c) distribution of energy inside these two devices.

This result indicates that way over half of the energy emitted in the conventional VCSEL is contained in the form of coherent oscillations of the bound (valence) electrons, which is exactly what is being professed as a unique property of the spaser emission. It is only when the energy exits the laser cavity that it is fully transferred to the electro-magnetic field (the mismatch between the two forms of energy is precisely what causes the Fresnel reflection). But of course, it would not be all that difficult to have nearly 100% reflective mirrors and large intrinsic loss (just as in spaser) and insure that no photons would ever leave the VCSEL cavity. Whether this would be desirable is a different question. But if one claims that the main and unique property of SPASER is the nature of the emitted quanta being mostly charge oscillation, then any ubiquitous semiconductor lasers have the same properties.

For the sake of comparison let us estimate the same parameter $f_{field}$ for the sub-wavelength spaser shown generically in Fig. 1b. Since the mode is sub-wavelength the magnetic field is very weak and its energy, roughly $\varsigma\varepsilon_r E^2/4$, where $\varsigma = (\pi an/\lambda)^2$ and $a$ is the mode diameter, is smaller than electrical energy. The energy conservation is preserved by the kinetic energy of free electrons $\partial(\chi_m\omega)/\partial\omega$, where $\chi_m$ is the free electron susceptibility. Since the spaser requires high gain it operates far above the bandgap in semiconductor and the dispersion there is small, hence

$$f_{field}^{SPASER} = \frac{1+\varsigma\varepsilon_r}{2\varepsilon_r} \approx 0.21. \tag{3}$$

The difference between (2) and (3) as illustrated by the diagrams in Fig. 1c is not that large at all. **In both cases more than 50% of the energy inside the operating mode is contained in the form of charge oscillations** – the main difference is that for spaser it is the charge of free electrons in the metal while for VCSEL it is the bound charge of electrons in the valence band of semiconductor. In both cases once the energy that leaves the mode it is in the form of photons, i.e. quanta of electro-magnetic field, but, since the charge oscillations in the metal decay at the rate of $10^{14}$ s$^{-1}$ [3,4] and the radiative efficiency of small nanoparticle is not so high, most of the energy in spaser dissipates inside the metal rather than exiting. Whether this can be considered an advantage of spaser or not is an open question, but, of course, one can easily introduce large cavity loss into the VCSEL and mimic the spaser performance in the sense that no photons ever emerge from the device, if the need for such contraption ever arises.

**Spaser vs. VCSEL: threshold, input-output characteristics, and linewidth**

Next we compare the power and coherence characteristics of VCSEL and spaser, both of which can be described by a set of coupled rate equations [16].

$$\frac{dn_c}{dt} = \frac{I}{e} - \frac{1}{\tau_s}\left(\gamma^{-1}n_p + \beta^{-1}\right)n_c$$
$$\frac{dn_p}{dt} = \frac{1}{\tau_s}\left(\gamma^{-1}n_p + 1\right)n_c - \frac{n_p}{\tau_c} \tag{1}$$

where $n_c$ is the total number of electron-hole pairs in the active region and $n_p$ is the total number of polaritons in the cavity. Following the discussion in the previous section, we use the term **"polariton"** rather than **"photon"** because in any semiconductor laser the emitted quasi-particle of energy $\hbar\omega$ is a combination of a photon and the coherent oscillations of electrons inside the valence band, hence polariton. In case of a spaser the quasi-particles also include contribution of oscillations of free electron in the metal so it is a plasmon-polariton, but the physics remains the same. Other parameters in (1) are the pump current $I$, polariton lifetime in the cavity $\tau_c$, the spontaneous emission rate into the lasing mode $\tau_s^{-1}$ related to the total recombination rate $\tau_r^{-1}$ via the spontaneous emission factor $\beta = \tau_r / \tau_s \leq 1$, and the excess noise factor $\gamma \geq 1$ due to the non-zero population of the lower lasing state and for the semiconductor gain media equal to $\gamma \sim f_c(1-f_v)/(f_c - f_v)$ where $f_c$ and $f_v$ are the Fermi-Dirac function in the conduction and valence bands respectively. The main difference between VCSEL and spaser is in the factor of $\beta$ which is much smaller than 1 for VCSEL (which can emit into a number of confined and free space modes) and is close to 1 for a sub-wavelength spaser with only a single mode overlapping with the gain region in both spectral and spatial domains.

We now introduce the normalized variables, $N_c = n_c \tau_c / \gamma \tau_s$, $N_p = \gamma^{-1} \beta n_p$, and $P = \tau_c I / e\gamma(1+\beta^{-1})$ and re-write (1) as

$$\frac{dN_c}{dt} = \frac{1}{\tau_r}\left[P(1+\beta) - N_c(N_p + 1)\right]$$
$$\frac{dN_p}{dt} = \frac{1}{\tau_c}\left[N_c(N_p + \beta) - N_p\right]. \qquad (2)$$

The steady-state solution of (2) is

$$N_c = \frac{N_p}{N_p + \beta}; \; P = \frac{N_p}{N_p + \beta}\frac{N_p + 1}{\beta + 1}. \qquad (3)$$

Clearly, for the typical laser in which the spontaneous emission is either coupled to a large number of modes, or is insignificant compared to non-radiative decay $\beta \ll 1$ and one obtains carrier density clamped at the threshold value $N_c = 1$, and the number of photons being equal to roughly $N_p \approx P - 1$ i.e. input-output characteristics shows a typical threshold bending point near $P = 1$ corresponding to the threshold current $I_t = e\gamma/\beta\tau_c$. If we estimate the total power dissipation due to radiative and nonradiative losses of the cavity we obtain $W = \hbar\omega n_p / \tau_c = (\hbar\omega/e)(I - I_t)$.

At the opposite extreme is a single mode laser (or spaser) with $\beta \approx 1$ for which we obtain $N_c = 1/(1+N_p^{-1})$, i.e. the carrier density does not clamp at the threshold but slowly saturates at the same value of $N_c = 1$ and the output characteristics becomes thresholdless $N_p = 2P$. If we estimate the total power dissipation due to radiative and nonradiative losses of the cavity we obtain the obvious result $W = \hbar\omega n_p / \tau_c = (\hbar\omega/e)I$ indicating that all the energy of the injected carriers (save the quantum defect between the voltage and polariton energy) ends up transformed into polaritons in the cavity.

Now, since the input-output characteristic exhibits no kink the threshold can be defined by the second order auto-correlation function assuming value of unity, which is the same as reduction of the linewidth by a factor of two, as has been shown in [15]. The linewidth can be found from (2) as

$$\Delta\omega = \frac{1}{\tau_c}\frac{\beta}{N_p + \beta} = \frac{1}{\tau_c}\frac{\gamma}{n_p + \gamma} \tag{4}$$

and hence the threshold value of the normalized polariton number is $N_{p,t} = \beta$, i.e., the actual photon number is $n_{p,th} = \gamma$ which is open to a very simple interpretation – for each coherent polariton emitted there is also one incoherent polariton emitted spontaneously. The threshold value of pump is then according to the second equation in (3) is $P_t = 1/2$ and the value of threshold current is $I_t^{'} = e\gamma(\beta+1)/2\beta\tau_c$ i.e. this definition differs from the one obtained above by a factor ranging from 0.5 to 1 ( [17]).

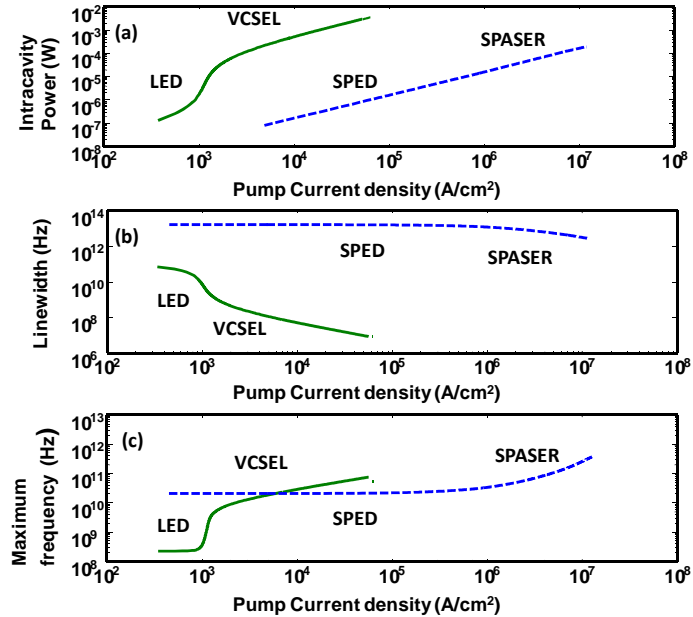

**Figure 2.** (a) Input-Output Characteristics of VCSEL and Spaser (b) Linewidth vs. pump current density of VCSEL and Spaser (c) Maximum operational frequencies of VCSEL and Spaser

If one analyzes a typical VCSEL with about 5 mm mode size and cavity lifetime $\tau_c \approx 3ps$ one gets the value of $\beta \approx 10^{-3}$ and $\gamma \approx 4$ which makes the threshold current about 0.1mA, in comparison, experimental results are a few times higher due to non-ideal effects such as poor current confinement. In terms of current density it corresponds to about 300A/cm², as shown in the input-output characteristics in Fig. 2a. If one now compares this result with a single mode spaser with cavity lifetime determined by the rate of loss in metal, i.e., optimistically speaking $\tau_c \approx 10 fs$, $\beta \approx 1$ and $\gamma \approx 2$ (the latter due to higher position of quasi-Fermi levels inside the bands) one arrives at the threshold current, as expected, about 30µA, but for the roughly 30nm radius plasmonic mode it would correspond to the current density of no less than 1MA/cm². This characteristic is also shown as dashed curve in Fig. 2a.

The threshold current decreases with the increase in $\beta$, i.e. with reduction in the number of modes coupled with gain, hence the desire to reduce the volume of the laser. At the same time the reduction of the volume usually implies the reduction of the photon lifetime, but still reducing the dimensions usually pays off up to the point where diffraction losses increase dramatically and reduce cavity time $\tau_c$. This is the main rationale for reducing the volume of the cavity and the ultra-low thresholds, as low as 4µA [18] have been obtained for small lasers based on photonic crystal cavities. But once the volume is being reduced beyond $(\lambda/2n)^3$ the number of modes remains the same, just one, no matter how small is the cavity and the threshold current does not change, while the threshold current density continues to increase.

Consider now the evolution of linewidth (4) with power for the typical laser with $\beta \ll 1$ linewidth quickly collapses above the threshold as

$$\Delta\omega \sim \beta\tau_c^{-1}(P-1)^{-1} = \beta\tau_c^{-1}(I/I_t - 1)^{-1} \qquad (5)$$

which is of course the Townes-Schawlow linewidth (obviously (5) is applicable only above the lasing threshold). As shown in the solid curve of Fig. 2b, the linewidth of the VCSEL that is pumped 10 times above the threshold, i.e. 3kA/cm², becomes about 10MHz which is less than experimentally obtained results [19] once again by a factor of a few due to non-ideal effects. But the goal here is to obtain just order-of-magnitude estimates.

But for a single mode laser (spaser) the linewidth narrowing is far more gradual as we obtain

$$\Delta\omega \sim \tau_c^{-1}(2P+1)^{-1} = \tau_c^{-1}(I/I_t + 1)^{-1}. \qquad (6)$$

This result which is characterized by the absence of abrupt change at the threshold is due to the fact that in general the linewidth can be reduced only by a relatively small factor compared to the linewidth of the unpumped (or "cold") cavity. Since for the subwavelength cavities incorporating the losses are high ($\tau_c \sim 10 fs$) it appears that the linewidth of the laser will remain very wide (dashed line in Fig. 2b).

Even if we pump the spaser 10 times above threshold to 10 MA/cm² the linewidth will still be on the scale of 1THz indicating that spaser will never achieve degree of coherence required, for instance, to transmit information in the coherent format. This is a consequence of two inherent properties of the plasmonic nanocavity – it is small, hence all the spontaneous emission ends up in just one mode, and its losses are very high which requires this spontaneous recombination to be very fast (which is due to Purcell enhancement). Thus enormous amounts of both amplitude noise and phase noise are introduced, the latter of which is responsible for the broad linewidth.

**Spaser, SPED, and VCSEL – who is faster?**

This leaves us with one last potential advantage for a spaser, but a very important one – that is the fast speed. The lasers usually exhibit much faster speed than spontaneous sources, such as LED's since the relatively slow spontaneous recombination process gets enhanced by the stimulated recombination. Let us see whether this is true for a spaser in which the number of photons (polaritons) inside the mode cannot be high. To do so we consider the small signal modulation, i.e. each variable will have CW bias point value (3) and the harmonic signal of frequency $\Omega$ - $P(t) = P + \tilde{P}e^{-j\Omega t}$, $N_c(t) = N_c + \tilde{N}_c e^{-j\Omega t}$, and $N_p(t) = N_p + \tilde{N}_p e^{-j\Omega t}$, and substitute it into (2) to obtain the set of coupled equations

$$-j\Omega\tau_r \tilde{N}_c = \tilde{P}(1+\beta) - (N_p+1)\tilde{N}_c - N_c \tilde{N}_p$$
$$-j\Omega\tau_c \tilde{N}_p = (N_p+\beta)\tilde{N}_c + (N_c-1)\tilde{N}_p \tag{7}$$

which has this solution for the photon density

$$\tilde{N}_p(\Omega) = \frac{-A(\beta)\tilde{P}\Omega_0^2}{\Omega_0^2 - \Omega^2 - j\Omega\Gamma} \tag{8}$$

where the resonance frequency of relaxation oscillation is

$$\Omega_0 = \frac{1}{\sqrt{\tau_c \tau_r}} \left[ N_p + \beta + \frac{\beta(1-\beta)}{N_p+\beta} \right]^{1/2} \approx \begin{cases} \sqrt{\dfrac{N_p}{\tau_c \tau_r}} & \beta \ll 1 \\ \sqrt{\dfrac{N_p+1}{\tau_c \tau_r}} & \beta = 1 \end{cases}, \tag{9}$$

the damping constant is

$$\Gamma = \frac{N_p+1}{\tau_r} + \frac{\beta}{N_p+\beta}\frac{1}{\tau_c} = \begin{cases} \dfrac{N_p+1}{\tau_r} & \beta \ll 1 \\ \dfrac{N_p+1}{\tau_r} + \dfrac{1}{N_p+1}\dfrac{1}{\tau_c} & \beta = 1 \end{cases}, \tag{10}$$

and the coefficient in front $A(\beta)$ varies between 1 for $\beta \ll 1$ and 2 for $\beta = 1$.

As one can see the resonant relaxation frequency in the single mode spaser differs from the resonant frequency of a conventional laser with many modes by one extra photon, which is not a significant difference. The damping constant, on the other hand, is much larger for the single mode spaser because the polariton decay is much faster than the recombination time, even if the latter is enhanced by the Purcell effect. Therefore, for the spaser it is expected that $\Gamma \gg \Omega_0$, i.e. the spaser response remains over-damped well beyond the threshold in contrast to the VCSEL that exhibits standard under-damped response.

From the frequency response it is not that difficult to determine a maximum frequency $f_{max}$, which we define here as a frequency at which the response to modulation (8) decreases by a factor of $2^{-1/2}$ relative to DC. The results are shown in Fig. 3c – the solid curve for VCSEL immediately demonstrates why semiconductor lasers operating above threshold have far superior frequency responses than LED's – as one can see near the threshold $f_{max}$ jumps from a 250MHz to a few GHz and then ramps up to the value of roughly 50GHz at pump current density of $6 \times 10^4$ A/cm$^2$. But for the spaser (the dashed line), the situation is different – even below threshold operating in the SPED regime the response is fast, approaching 30GHz due to the Purcell enhancement then it starts gradually increasing reaching about 120GHz at threshold and only at very high current density of $10^7$ A/cm$^2$ it reaches 500 GHz. The difference between SPED and spaser is not nearly as dramatic as the difference between a LED and a VSCSEL. It follows from the simple fact that with very few polaritons in the spaser mode the electron and photon populations remain weakly coupled – hence the speed of the device is determined by the slower process of carrier recombination rather than by the faster process of photon escape from the mode.

One should also note that for the sake of simplicity we have considered the gain $g(n_c)$ to be a linear function of carrier density $n_c$. By doing so we have neglected an important effect of reduction in the differential gain of semiconductor laser that happens when the Fermi levels are deep inside the bands, which is, of course, precisely what happens when the carrier density required to compensate for the metal loss in SPASER approaches $10^{19}$ cm$^{-3}$. Essentially the expression for resonant frequency (9) must be multiplied by a factor $[\partial g(n_c)/\partial n_c]^{1/2} [g(n_c)/n_c]^{-1/2}$ [16] which is sure to decrease the maximum frequency of a spaser by a significant amount. Furthermore, we have not considered the gain compression phenomena [19] that would be especially prevalent in the small spaser cavity with high density of electric field and thus bound to depress the spaser modulation response even further.

**Discussion and conclusions**

Our study has yielded these important points

(1) When it comes to the nature of what the emitted quanta are, the spaser is not fundamentally different from any existing solid state laser – in both cases the emitted quasi-particles are polaritons, i.e. combinations of photons and electronic oscillations. In case of solid state laser, especially a semiconductor laser (VCSEL) the electronic oscillations are those of bound electrons, while for spaser it is the oscillations of free electrons in the metal. For VCSEL operating near the band edge of semiconductor up to 60% of energy is contained in the oscillations of bound electrons. For spaser this value can be as high as 80%, but that is a quantitative and not a fundamental difference.

(2) Since the emitted energy in spaser is contained inside the inherently lossy oscillations of free electrons in metal, and since all the energy is channeled into the single mode the threshold of spaser (defined as the current density corresponding to linewidth reduction by a factor of two) is very high and determined only by the metal loss. The current density in excess of $1MA/cm^2$ is difficult to achieve, especially when it comes to the transport of holes. Even in absolute numbers of current, the threshold of spaser is on the order of 50 µA while for the 3 mm diameter VCSEL it is only three times as high, and, as shown the results for photonic crystal microcavity laser can be reduced to just a few µA [18]. Since the out-coupling efficiency of VCSEL is much higher, one can achieve true useful output power (rather than the power dissipated in the metal) with VCSEL which is orders of magnitude higher than that of spaser with only a few times increase in input power.

(3) The single mode nature of spaser and its high metal loss mean that the coherence properties are very poor no matter what and there is no dramatic improvement once the spaser goes over the threshold. Even when pumped to probably unsustainable power densities in excess of $10MA/cm^2$ the linewidth of SPASER is in the range of 3THz, while for VCSEL the linewidth as narrow as a few tens of MHz is attainable.

(4) Finally, the most touted advantage of the spaser – fast modulation speed is not obvious. Since the count of polaritons inside the small lossy cavity is always low, the speed of spaser is not much higher than that of SPED, i.e. the recombination rate enhanced by the Purcell effect. Only at extremely high, and probably unsustainable, current densities of $10 MA/cm^2$ one can obtain modulation speeds of 100's of GHz, near the threshold, however, the performance only begins to approach 100GHz which is what can be realistically obtained with VCSEL.

It should be noted that for the spaser we have considered the best case scenario. The spaser was assumed to operate under ideal (and not very likely) conditions (no quenching by high order plasmonic modes, no current shortening via the metal rather than semiconductor junction, no reduction in differential gain, and no gain compression). It is difficult to predict whether such conditions can be attained in practice, because no sub-wavelength electrically pumped spaser has ever been operational. At the same time the results for VCSEL are realistic as they agree well with the wealth of experimental data.

In conclusion, we have performed comparative analysis between currently available VCSEL's and proposed sources based on sub-wavelength confinement assisted by plasmons – both coherent (spaser) and incoherent (SPED). Based on the results, reported here, the researchers can make an informed judgment about which of the devices would be most suitable for each particular application.

**Methods**

The analysis conducted in this comparative work was based on a combination of analytical derivations and numerical simulations. Properties associated with VECSELs are derived from the well-established theory on semiconductor lasers with GaAs as the active medium. For spaser and SPED, we have considered silver nanoparticles in the shape of prolate spheroid surrounded by GaAs as the gain medium in order to allow for tuning of the SP resonance by the aspect ratio of the silver spheroid. The SP modes were obtained using the finite-difference time-domain (FDTD) method commercially available from Lumerical, Inc. Characteristics for spaser and SPED were obtained analytically and the details of the theoretical work has been published previously [11].

**Author contributions**

J B K performed analytical derivations and G S numerical analysis.